\documentclass[11pt,onecolumn,a4paper,dvips]{IEEEtran}

\usepackage{mathrsfs}
\usepackage{amsfonts}
\usepackage{amssymb}
\usepackage{stfloats}
\usepackage{cite}
\usepackage{graphicx}
\usepackage{psfrag}
\usepackage{subfigure}
\usepackage{amsmath}
\usepackage{array}

\usepackage{bm}
\usepackage{algorithm}

\usepackage{color}

\newtheorem{Thm}{Theorem}
\newtheorem{Lem}{Lemma}

\newtheorem{Def}{Definition}

\newtheorem{Rem}{Remark}

\title{Game Theoretical Power Control for Open-Loop Overlaid  Network MIMO Systems with Partial Cooperation}

\author{Hao~Yu,~\IEEEmembership{Student~Member,~IEEE,}
        Shunqing Zhang,~\IEEEmembership{Member,~IEEE,}
        Vincent~K.~N.~Lau,~\IEEEmembership{Senior~Member,~IEEE,}% <-this % stops a space
\thanks{Hao Yu and Vincent K. N. Lau are with the Department of Electronic and Computer Engineering
(ECE), The Hong Kong University of Science and Technology (HKUST),~Hong Kong (e-mails:
yuhao@ust.hk; eeknlau@ust.hk).}% <-this % stops a space
\thanks{Shunqing Zhang was with the Department of Electronic and
Computer Engineering (ECE), The Hong Kong University of Science and Technology
(HKUST),~Hong Kong. He is currently with Huawei Technologies, ~Shanghai, ~China (e-mail:
sqzhang@huawei.com).}  }
\begin{document}

\markboth{IEEE TRANSACTION ON WIRELESS COMMUNICATIONS, ACCEPTED} {Shell
\MakeLowercase{\textit{et al.}}: Bare Demo of IEEEtran.cls for Journals}

\maketitle

\begin{abstract}
Network MIMO is considered to be a key solution for the next generation wireless systems
in breaking the interference bottleneck in cellular systems. In the MIMO systems,
open-loop transmission scheme is used to support mobile stations (MSs) with high
mobilities because the base stations (BSs) do not need to track the fast varying channel
fading. In this paper, we consider an open-loop network MIMO system with $K$ BSs serving
K {\em private MSs} and $M^c$ {\em common MS} based on a novel {\em partial cooperation}
overlaying scheme. Exploiting the heterogeneous path gains between the private MSs and
the common MSs, each of the $K$ BSs serves a private MS non-cooperatively and the $K$ BSs
also serve the $M^c$ common MSs cooperatively. The proposed scheme does not require
closed loop instantaneous channel state information feedback, which is highly desirable
for high mobility users. Furthermore, we formulate the long-term distributive power
allocation problem between the private MSs and the common MSs at each of the $K$ BSs
using a {\em partial cooperative} game. We show that the long-term power allocation game
has a unique Nash Equilibrium (NE) but standard best response update may not always
converge to the NE. As a result, we propose a low-complexity distributive long-term power
allocation algorithm which only relies on the local long-term channel statistics and has
provable convergence property. Through numerical simulations, we show that the proposed
open-loop SDMA scheme with long-term distributive power allocation can achieve
significant performance advantages over the other reference baseline schemes.
\end{abstract}


\IEEEpeerreviewmaketitle
\section{Introduction}
Inter-cell interference (ICI) has been widely considered as a critical performance
bottleneck for wireless communications in the cellular networks \cite{Blum03,Catreux}.
For instance, mobile stations (MSs) at the cell edge (within the coverage of multiple
base stations (BSs)) are usually interference limited. To alleviate the interference
issues, traditional cellular systems employ frequency reuse so as to control the
interference at the expense of poor spectral efficiency \cite{Cox82}. On the other hand,
network MIMO communications \cite{Kara,Gesbert} are considered to be a key solution for
the next generation wireless systems in breaking the interference bottleneck in cellular
systems. The idea of network MIMO communications is to utilize cooperation among multiple
BSs for joint signal processing in the uplink/downlink directions. Through cooperation,
the undesired ICI can be transformed into useful signals via the collaborative
transmission among adjacent BSs \cite{Jing,Shamai}. %Hence, network
%MIMO solution provides a fundamental solution to interference in
%cellular networks \cite{Gesbert}.

One key challenge in network MIMO systems is on how to spatially multiplex multiple MSs
effectively and efficiently. Traditionally, linear precoding (such as Tx-MMSE
\cite{Paulraj01} or zero-forcing \cite{Spencer04}) could be used to spatially multiplex
MSs but closed loop knowledge of instantaneous channel state information (CSI) is
required at the BS and this refers to closed-loop spatial division multiplexing access
(SDMA)\cite{Suard,Viswanath}. However, the closed-loop SDMA schemes only work for low
mobility MSs where the channel conditions remain quasi-static within the transmission
duration. For the high mobility MSs, it is very difficult to keep track of the channel
state information (CSI) at the BSs and hence, the above closed-loop schemes cannot be
applied for high mobility MSs.

Open-loop transmission scheme has been widely considered in the
existing literature for high mobility users. By open-loop schemes,
we mean that instantaneous CSI knowledge is not required at the BS.
 For example, in \cite{Tarokh,Tarokh_diversity}, the authors
proposed an open-loop transmission scheme, namely the space-time
block code (STBC) for point-to-point scenarios. In \cite{DSTTD},
double-STTD which is able to serve two users simultaneously has been
proposed for 4 transmit antenna and 2 receive antenna MIMO link to
fully exploit the spatial diversity and spatial multiplexing gains.
All the above open-loop schemes only work in a single cell scenario
and there are several important technical challenges to extend to
multi-cell systems. They are elaborated below.

\begin{itemize}
\item{\bf Heterogeneous Path Gain and Shadowing Effect}  In the network MIMO systems, the path gains of
different MSs are quite different. The heterogeneous path gain
effect for different types of MSs leads to significant power
efficiency loss and hence the conventional open-loop transmission
schemes (which have ignored the heterogeneous path gain and
shadowing effects)
cannot be directly applied in the network MIMO. %For example, if we
%serve different types of MSs in different time slots with a certain
%target transmission rate requirement, the power requirement for the
%MS with a large path gain is much less than the MS with a small path
%gain. As a result, the transmit power at each BS cannot be fully
%utilized when serving the MS with a large path gain.

\item{\bf Dynamic and Heterogeneous MIMO Configurations}
In the network MIMO systems, the number of cooperating BSs is
changing dynamically and hence, we are not able to use existing STBC
structures in such dynamic MIMO configurations (with time varying
number of transmit antennas). %One possibility is to apply identical
%STBC structure at each BS.  However, since the number of transmit
%antennas at different BSs may be variant, the STBC design for the
%cooperatively served MS has to accommodate for the worst case BS,
%i.e., the BS with minimum transmit antennas.
\end{itemize}

In this paper, we consider a network MIMO system with multiple BSs and multiple high
mobility MSs\footnote{As a result, the BS does not have knowledge of instantaneous
channel state information (CSI) of the MSs.}. We propose a novel open-loop scheme to
serve $K$ {\em private MSs} and $M^c$ {\em common MSs} simultaneously based on novel {\em
partial cooperative overlaying}. Specifically, each BS serves a private MS
non-cooperatively. By exploiting the path gain difference between the common MSs and the
private MS, the $K$ BSs also serve the $M^c$ common MSs cooperatively at the same
spectrum as the private MSs. The proposed scheme does not require knowledge of
instantaneous CSI at the BS and supports dynamic and flexible network MIMO
configurations. Furthermore, to adjust the {\em long-term power allocation} between the
common MSs and the private MS at each of the $K$ BSs, we formulate the {\em long-term
distributive power control} problem using a {\em partial cooperative game} formulation.
We show that the long-term power control game has a unique Nash Equilibrium (NE) but the
conventional {\em best response update} algorithm cannot always converge to the NE. As a
result, we propose a low-complexity distributive long-term power allocation algorithm
which only relies on the local channel statistics and has provable convergence property.
Through numerical simulations, we show that the proposed open-loop overlaying scheme with
a distributive long-term power allocation algorithm can achieve significant performance
advantages over the traditional schemes and the distributive algorithm has negligible
performance loss compared with the centralized power allocation scheme.

%The rest of the paper is organized as follows.
%Section~\ref{sect:chan_mod} introduces the network MIMO system
%model. We propose our open-loop SDMA scheme for network MIMO systems
%and formulate the joint long-term power control problem in
%Section~\ref{sect:prob_form}. The optimal long-term power scheduling
%algorithms are developed in Section~\ref{sect:sol}. In
%Section~\ref{sect:experiments}, we give some numerical results and
%discussions. Concluding remarks are given in
%Section~\ref{sect:conclusion}.

\subsection{Notations}
We adopt the following notation conventions. Boldface upper case letters denote matrices, boldface
lower case letters denote column vectors, and lightface italics denote scalers. $\mathbb{C}^{n
\times m}$ denotes the set of ${n \times m}$ matrices with complex-valued entries and the
superscript $(.)^H$ denotes Hermitian transpose operation. The matrix $\mathbf{I}_n$ denotes the
$n\times n$ identity matrix. Expressions $P_r(x)$ denotes the probability density function (p.d.f.)
of the random variable $x$. The expectation with respect to $x$ is written as
$\mathbb{E}_{x}[\cdot]$ or simply as $\mathbb{E}[\cdot]$.

\section{System Model}
\label{sect:chan_mod} 
\begin{figure}
\centering
\includegraphics[width = 4in]{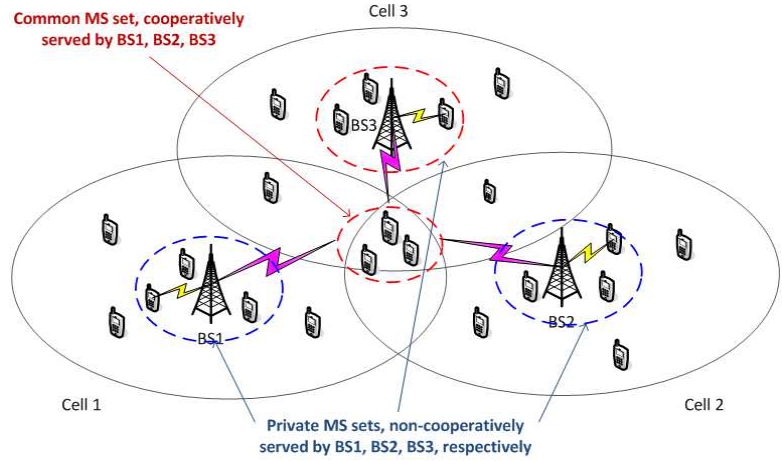}
\caption{An example of network configuration containing three cells. MSs in the solidline
red circle represent the common MS set which is served cooperatively by all the BSs
within coverage. MSs in the solidline blue circle represent the private MS sets which are
served non-cooperatively in the coverage of BS1, BS2, and BS3, respectively.}
\label{fig:sys_conf}
\end{figure}

Consider a cellular network where there are $K$ base stations (BSs)
and $M$ high mobility mobile stations (MSs) as shown in Figure \ref{fig:sys_conf}. We
assume each BS is equipped with $N_t$ transmit antennas\footnote{In this paper, we focus
on the case where all the BSs have the same number of antennas. As we elaborate later,
the proposed scheme can be directly applied to the dynamic and heterogeneous MIMO
configurations with little modification. } and each MS is equipped with $N_r$ receive
antennas.

Denote $T$ to be the transmission time intervals\footnote{Transmission time interval  is
defined to be the time duration where the channel fading coefficients in the multi-cell
network MIMO systems remain quasi-static.} and $\mathbf S_k \in \mathbb C^{N_t \times T}$
to be the transmitted signals from the $k$-th BS. The received signals of $m$-th MS,
denoted by $\mathbf{Y}_m \in \mathbb C^{N_r \times T}$, can thus be modeled as follows,
\begin{eqnarray}
\label{eqn:rev} \mathbf Y_{m}= \sum_{k=1}^{K}\sqrt{P_{k}L_{mk}}
\mathbf H_{mk} \mathbf S_{k} + \mathbf Z_{m}, \ \forall m = 1, 2,
\ldots,M
\end{eqnarray}
where $\mathbf{H}_{mk} \in \mathbb C^{N_r \times N_t}$ is the normalized complex fading
coefficients from the $k$-th BS to the $m$-th MS, $\mathbf Z_{m} \in \mathbb C^{N_r
\times T}$ is the additive white complex Gaussian noise (AWGN) with zero mean and unit
variances, $P_{k}$ denotes the transmit power of the $k$-th BS and $L_{mk}$ denotes the
long-term path gain and shadowing from the $k$-th BS to the $m$-th MS.

The following assumptions are made through the rest of the paper.
Firstly, all the receivers in the system have perfect CSI of each
corresponding link, i.e. the $l$-th MS has the perfect CSI knowledge
from the $k$-th BS. Secondly, we assume all the BSs have no
instantaneous CSI knowledge $\{\mathbf{H}_{mk}, m=1,2,\ldots,M,
k=1,2,\ldots,K \}$. Thirdly, all the entries of the channel
coefficient matrix $\{\mathbf{H}_{mk}, m=1,2,\ldots,M,
k=1,2,\ldots,K\}$ are independent and identical distributed (i.i.d.)
complex Gaussian random variables with zero mean and unit variance.
Moreover, we consider block fading channels where the aggregate CSI
$\mathbf H = \{\mathbf{H}_{mk}, m=1,2,\ldots,M, k=1,2,\ldots,K\}$
remains quasi-static within a fading block (i.e. the transmission
time interval $T$) but varies between different fading blocks.

\section{Problem Formulation}
\label{sect:prob_form} In this section we shall first introduce a user scheduling
algorithm, which classifies the high mobility MSs into $K$ {\em private MS sets} (one for
each BS) and a {\em common MS set} (shared by all the BS). Based on the user scheduling
algorithm, we propose a novel open loop scheme to overlay the MSs in the common set and
the private sets simultaneously using partial cooperation. We shall then discuss the
problem formulation of the long-term power allocation control in what follows.

\subsection{Long-term User Scheduling Algorithm}
We first define the private MS set and the common MS set below.

\begin{Def}[Common/Private MS Sets]~

\begin{itemize}
\item {\bf $k$-th Private MS Set:} The $k$-th {\em private MS set} $\mathcal{U}_k^p$ consists of one MS (the $m$-th MS) in which the long-term path gain and shadowing configuration $\{L_{m1}, L_{m2}, \ldots, L_{mK}\}$ satisfies the following criteria:
$L_{mk} - L_{mj} > \xi_{k}^{p}, \forall j \neq k$, where $\xi_{k}^{p}$ is the {\em $k$-th
private MS set threshold}.

\item {\bf Common MS Set:} The {\em common MS set} $\mathcal{U}^c$ consists of at most $M^c$ MSs such that
$|L_{mk}-\frac{1}{K}\sum_{j=1}^{K}L_{mj}|\leq \xi^{c}, \forall k=1,2,\ldots,K$ for all $m
\in \mathbf{U}^c$, where $\xi^{c}$ is the {\em common MS set threshold}.
\end{itemize}\label{def:MS_Set}
\end{Def}

\begin{Rem}
To ensure $\mathcal{U}^c \bigcap \mathcal{U}^p_k = \varnothing$ for all $k=1,\ldots,K$,
the thresholds need to satisfy $\xi_k^p \geq \frac{K-1}{K}\xi^c$ for all $k=1, \ldots,
K$. As such, the private MS sets consist of MSs closer to the home cell whereas the
common MS set consists of the MSs closer to the "coverage overlap areas" between the BSs.
\end{Rem}
\begin{Rem}
On the other hand, a MS may belong to neither of the above two set. In that case, the MS
is not selected to be the "common MS" or the "private MS" and does not participate in the
"open loop overlaying scheme". This MS may be served in the normal way (e.g. assigned
another sub-band). Since the MS are moving around, this particular MS may be able to be
selected as the "common MS" or "private MS" in some future time.
\end{Rem}

Algorithm 1 illustrates a low complexity user scheduling
algorithm to construct $\mathcal{U}^p_k, k=1, \ldots, K$ and $\mathcal{U}^c$ based on the
{\em local long-term path gain and shadowing} at each of the MSs.
\begin{algorithm}
\caption{Long-term User Scheduling Algorithm} \label{alg1:User_Scheduling}
\begin{itemize}
\item \textbf{Step 1: MS Broadcast:}

At the $m$-th ($m=1,2,\dots,M$) MS side, the $m$-th MS measures path gains $L_{mk},
k=1,\ldots,K$(in dB) from all the $K$ BSs. According to Definition \ref{def:MS_Set}, if
there exists $L_{mk}$ such that $L_{mk}-L_{mj} > \xi_{k}^{p}, \forall j\neq k$, then the
$m$-th MS labels itself as a potential member of the $k$th private MS set; if $\forall
k=1,2,\ldots,K$, $|L_{mj}-\frac{1}{K}\sum_{j=1}^{K}L_{mj}|\leq \xi^{c}$, then the $m$-th
MS labels itself as a potential member of the common MS set. Those MS being potential
members of the private set or the common set will then broadcasts its (label,BS$\_$ID) to
all the $K$ BSs.

\item \textbf{Step 2: Formation of the Private MS Set:}

Denote $\mathcal{U}_{k}^{p}, k=1,2,\dots,K$ to be the $k$-th private MS set. The k-th BS
picks one MS with label = "private" and BS$\_$ID = $k$ to be the member of the private MS
set $\mathcal{U}_{k}^{p}, k=1,2,\dots,K$ randomly.

\item \textbf{Step 3: Formation of the common MS Set:
}

Denote $\mathcal{U}^{c}_k$ to be the potential common MS set at the $k$-th BS. Assign the
$m$-th MS to $\mathcal{U}^{c}_k$ at the $k$-th BS if the label from the $m$-th MS is
"COMMON". Each BS then submits $\mathcal{U}_k^c$ to the base station controller (BSC). At
the BSC, the common MS set $\mathcal{U}^{c}$ is chosen as an intersection of
$\mathcal{U}_{k}^{c}, k=1,2,\dots,K$, i.e. $\bigcap_{k=1}^{K}\mathcal{U}_{k}^{c}$. If the
number of members in the intersection exceeds $M^c$, then $M^c$ users will be selected
randomly.

\end{itemize}
\end{algorithm}

\subsection{Signal Model for the Private/Common MS}
Given the user sets $\mathcal U^p_k, k=1, \ldots, K$ and $\mathcal U^c$, the received
signal of the $m$-th MS in the private and common MS sets, denoted by $\mathbf Y_{m}$ ,
is given by:
\begin{eqnarray} \label{eqn:rev_user_scheduing}
\mathbf Y_{m} = \left\{ \begin{array}{ll}
\sum_{k =1}^{K}\sqrt{P_{k}L_{mk}} \mathbf H_{mk} \mathbf S_{k} +\mathbf Z_{m},  &  \qquad m \in \mathcal{U}^c \\
\sqrt{P_{k}L_{mk}} \mathbf H_{mk} \mathbf S_{k} + \mathbf Z_{m}, &
\qquad m \in \mathcal{U}^{p}_{k}
\end{array} \right.
\end{eqnarray}
\begin{Rem}
At the private MS, inter-cell interference doesn't appear in the received signal model
because the inter-cell interference at the private MS is very weak and negligible. For
instance, if $\xi_{k}^{p} = 20$dB, $k=1,2,\ldots,K$, the inter-cell interference would be
100 times less than the useful signal.
\end{Rem}

\subsection{Open-Loop Overlaying
Transmission/Dection Scheme}
\subsubsection{Open-Loop Overlaying Transmission Scheme}
\begin{figure}
\centering
\includegraphics[width = 5in]{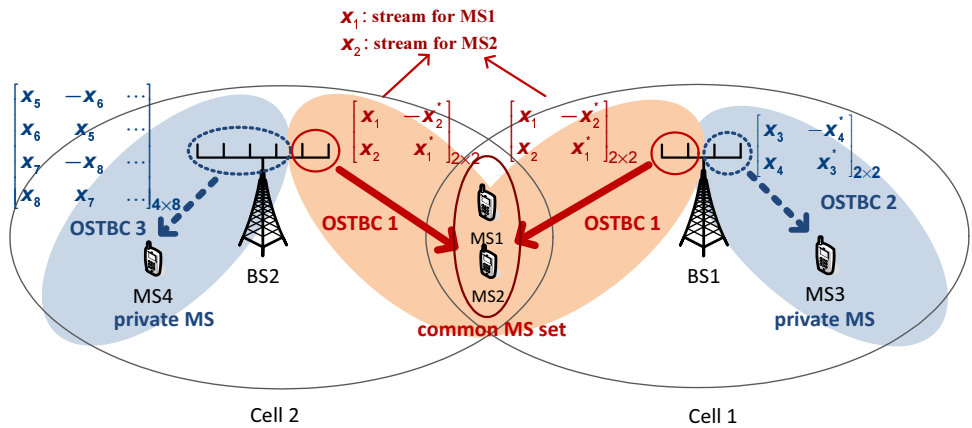}
\caption{An illustrative diagram of the transmit structure at each BS. For example, the
BS1 and BS2 are serving the common set (MS1 and MS2) cooperatively with the same OSTBC
(in which stream x1 is for MS1 and stream x2 is for MS2) and serving the private MS3 and
MS4 using two different OSTBC structures respectively.} \label{fig:trans}
\end{figure}

Consider the information streams for the $M^c$ MSs in the common MS set (denoted by
$\mathbf X_{j}, j\in\mathcal{U}^c$) and the information streams for the $l$-th MS (in the
$k$-th private MS set) (denoted by $\mathbf X_l, l\in\mathcal{U}^p_k$) are transmitted
over the $N_t$ antennas at the $k$-th BS. To exploit the possible diversity provided by
the transmit antenna arrays, orthogonal space-time block code (OSTBC)
\cite{Alamouti,Tarokh} scheme is applied for transmission, which spans over the entire
transmitting antennas. The information streams ($\mathbf X_{l}, l\in\mathcal{U}^c$) for
the $M^c$ common MSs\footnote{$M^c$ is limited to be less or equal to the number of
streams in OSTBC $\mathbf S^{c}$.} are jointly encoded as the OSTBC $\mathbf S^{c}$ and
the information streams for the MS in the $k$-th private MS set is encoded as the OSTBC
$\mathbf S^{p}_k$ as shown in Fig. \ref{fig:trans}. Without loss of generality, we assume
at the $k$-th BS, the two OSTBCs $\mathbf S^{p}_k$ and $\mathbf S^{c}$ are delivered
through $N_t^{p}$ and $N_t^{c}$ transmit antennas respectively with $N_t^{p} + N_t^{c} =
N_t$. So the transmitted symbols at the $k$-th BS are given by $\mathbf S_k = \left[
(\mathbf S^{p}_k)^{T} \ (\mathbf S^{c})^{T} \right]^{T}$.

For illustration purpose, let us consider a specific case with $M^c = 2$. Assume $N_t =
4$ with $N_t^{p} = N_t^{c} = 2$. At each BS, the two information streams for two MSs in
the common MS set respectively are jointly OSTBC encoded into one Alamouti's structure
\cite{Alamouti} and the information streams for the $k$-th private MS set are encoded as
the other Alamouti structure. The whole transmit structure is also known as double
space-time transmit diversity (D-STTD) \cite{DSTTD}. The transmitted structure at the
$k$-th BS is given by:
\begin{equation} \label{tran_signal}
\mathbf S_k=\sqrt{\frac{\theta_{k}^{p}}{N_{t}^{p}}} \left[\begin{array}{cc} s_{k,1}^{p} & -s_{k,2}^{p,*} \\ s_{k,2}^{p} & s_{k,1}^{p,*} \\ 0 &0 \\
0 &0\end{array} \right ]+\sqrt{\frac{\theta_{k}^{c}}{N_{t}^{c}}}\left [\begin{array}{cc}  0 &0 \\
0 &0 \\ s_{1}^{c} & - s_{2}^{c,*} \\ s_{2}^{c} &
s_{1}^{c,*}\end{array} \right ]
\end{equation}
where $\mathbf S^{p}_k = \left[
\begin{array}{cc}
s_{k,1}^{p} & - s_{k,2}^{p,*} \\
s_{k,2}^{p} &  s_{k,1}^{p,*}
\end{array}
\right]$ and $\mathbf S^{c} = \left[
\begin{array}{cc}
s_{1}^{c} & - s_{2}^{c,*} \\
s_{2}^{c} &  s_{1}^{c,*}
\end{array}
\right]$, $\theta_k^p $ is the power allocation ratio for the private MS set in the
coverage of the $k$-th BS and $\theta_k^c $ is the power allocation ratio for the common
MS set at the $k$-th BS. $\theta_k^p$ and $\theta_k^c$ satisfy the relation $\theta_k^p +
\theta_k^c =1$.

The proposed open-loop oveylaying scheme has the following
advantages.
\begin{itemize}
\item{\em Exploiting the Heterogeneous Path Gain:}
As we have mentioned before, the heterogeneous path gain effect for different types of
MSs leads to significant power efficiency loss and hence the conventional open-loop
transmission schemes cannot be directly applied in the network MIMO system. With the
proposed open-loop overlaying scheme, the common MS and the private MS can be
simultaneously served. Due to the long-term power splitting ratio $\theta_k^{c}$ and
$\theta_k^{p}$, we can efficiently control the ICI generated at the common MS side under
different path gain configurations through the carefully designed long-term power
allocation schemes to enhance the power efficiency for the private MS.

\item{\em Exploiting Flexible MIMO Configurations:}
In the proposed open-loop overlaying scheme, we could accommodate dynamic and
heterogenous MIMO configurations in the systems. This can be illustrated through the
following simple example. Consider three cooperative BSs with heterogeneous MIMO
configurations, e.g. BS1 is equipped with $4$ antennas, BS2 is equipped with $6$ antennas
and BS3 is equipped with $3$ antennas. Due to mobility of users, assume BS1 and BS2
cooperatively serve one common MS in the first time slot and BS2 and BS3 cooperatively
serve the common MS in the second time slot. In the traditional open-loop overlaying
scheme, the STBC design has to accommodate BS3 with three transmit antennas for both time
slots. However, for the proposed open-loop overlaying scheme as illustrated in Fig.
\ref{fig:trans}, BS1 and BS2 can use the remaining 2 and 4 transmit antennas for the
private MSs . In the second time slot, BS2 and BS3 can perform the similar operations to
serve the private MSs with the remaining transmit antennas. As a result, the proposed
open-loop overlaying scheme offers flexibility with respect to dynamic and heterogeneous
MIMO configurations in the systems.
\end{itemize}

\subsubsection{Open-loop Overlaying Detection Scheme}
 Applying the above transmission scheme, the
received signals at the common MS can be modeled as:
\begin{eqnarray}
\label{eqn:rev_edge_proof} \mathbf Y_m & = & \sum_{k=1}^{K}
\sqrt{P_{k}L_{mk}} \big[\mathbf{H}_{mk,1} \quad \mathbf{H}_{mk,2}
\big] \left[ \begin{array}{c}
\sqrt{\frac{\theta_{k}^{p}}{N_{t}^{p}}}\mathbf S^{p}_k \\
\sqrt{\frac{\theta_{k}^{c}}{N_{t}^{c}}}\mathbf
S^{c}  \end{array} \right] + \mathbf{Z}_m \notag \\
& = & \underbrace{
\sum_{k=1}^{K}\sqrt{\frac{P_{k}L_{mk}\theta_{k}^{c}}{N_{t}^{c}}}\mathbf{H}_{mk,2}
 \mathbf{S}^{c}}_{\textrm{Signal Part}} +
\underbrace{\sum_{k=1}^{K}
\sqrt{\frac{P_{k}L_{mk}\theta_{k}^{p}}{N_{t}^{p}}} \mathbf H_{mk,1}
\mathbf S_{k}^{p} + \mathbf Z_{m}}_{\textrm{Interference + Noise
Part}}, \qquad \forall m \in \mathcal{U}^c
\end{eqnarray}
Similarly, the received signals at the private MS can be modeled as:
\begin{eqnarray} \label{eqn:rev_center_proof} \mathbf Y_{m} & = &
\sqrt{P_{k}L_{mk}} \big[\mathbf{H}_{mk,1} \quad \mathbf{H}_{mk,2}
\big] \left[
\begin{array}{c}
\sqrt{\frac{\theta_{k}^{p}}{N_{t}^{p}}}\mathbf S^{p}_k \\
\sqrt{\frac{\theta_{k}^{c}}{N_{t}^{c}}}\mathbf
S^{c}  \end{array} \right] + \mathbf{Z}_{m} \notag \\
& = & \underbrace{\sqrt{\frac{P_{k}L_{mk}\theta_{k}^{p}}{N_{t}^{p}}}
\mathbf H_{mk,1}\mathbf S_{k}^{p}}_{\textrm{Signal Part}}+
\underbrace{\sqrt{\frac{P_{k}L_{mk}\theta_{k}^{c}}{N_{t}^{c}}}
\mathbf H_{mk,2}\mathbf S^{c} + \mathbf Z_{m}}_{\textrm{Interference
+ Noise Part}}, \qquad \forall m \in \mathcal{U}_k^{p}
\end{eqnarray}
where $\mathbf{H}_{mk,1} \in \mathbb C^{N_r \times N_{t}^{p}}$; $\mathbf{H}_{mk,2} \in
\mathbb C^{N_r \times N_{t}^{c}}$. $\mathbf{S}_{k}^{p}$ and $\mathbf{S}^{c}$ denote the
OSTBC encoded transmitted matrices for the private MS in the coverage of $k$-th BS and
the common MS set with entries $\pm s_{k,1}^{p}$, $\pm s_{k,1}^{p,*}$, $\ldots$, $\pm
s_{k,R^{p}_kT}^{p}$, $\pm s_{k,R^{p}_kT}^{p,*}$ and $\pm s_{k,1}^{c}$, $\pm
s_{k,1}^{c,*}$, $\ldots$, $\pm s_{k,R^{c}T}^{c}, \pm s_{k,R^{c}T}^{c, *}$ respectively.
$R_k^p$ is the encoding rate for the OSTBC $\mathbf S^p_k$ and $R^c$ is the encoding
rates for the OSTBC $\mathbf S^c$.

At the common MS side, the received signals are radio-frequency(RF)-combined to exploit
the {\em macro-diversity}. Based on \eqref{eqn:rev_edge_proof}, each MS in the common MS
set shall detect the whole OSTBC $\mathbf S^{c}$ by treating the interfering streams
$\mathbf S^{p}_{k}, k=1,2,\ldots,K$ as noise\footnote{Since the interfering streams are
contributed by the transmission to the private MSs, the power is much smaller due to the
heterogeneous path gain ($L_{jk} \ll L_{lk}, j \in \mathcal{U}^c, l \in \mathcal{U}^p_k$)
and hence, such detection scheme is reasonable for the weak interference scenarios
\cite{Tse,Blum03}.} and then take the desired stream from the decoded OSTBC $\mathbf
S^{c}$. At the private MS side, the received signal in \eqref{eqn:rev_center_proof}
corresponds to a "strong interference" scenario and hence, the MS in the $k$-th private
MS set shall first detect the interfering streams $\mathbf S^{c}$ and then perform
successive interference cancelation (SIC) to detect its own information streams $\mathbf
S_k^{p}$.

Using the OSTBC transmission structure and the above detection schemes, the throughput
expressions of the common MS and private MSs are summarized in the following lemma.

\begin{figure}
\centering
\includegraphics[width = 5in]{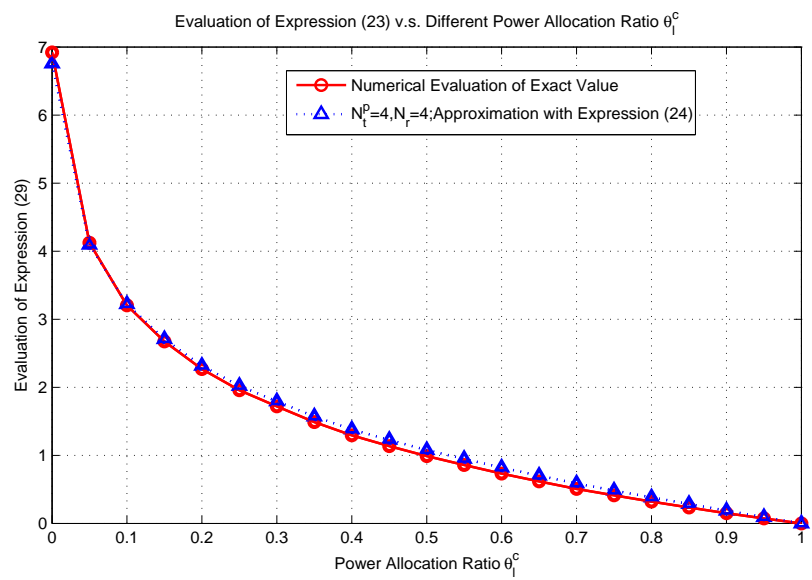}
\caption{Illustration of the quality of approximation in \eqref{eqn:appro0} and
\eqref{eqn:appro}.} \label{fig:appro}
\end{figure}

\begin{Lem}[Achievable Throughput]
\label{lem:throughput} Using the open-loop overlaying transmission and detection scheme
described above, the achievable throughput $\mathcal{C}_m$ of the MSs in the common and
private MS sets is given by:
\begin{eqnarray} \label{eqn:throughput}
\mathcal {C}_m  \approx \overline{\mathcal {C}}_m = \left \{ \begin{array}{l} \begin{array}{r}\frac{D_m}{D}\min_{j \in \mathcal{U}^c, l \in \mathcal{U}^p_k, k=1,\ldots,K}\{\overline{\mathcal{C}}_j^{c},\overline{\mathcal{C}}^{p}_l\}, \\  m \in \mathcal{U}^c \end{array}\\
\begin{array}{r}\log \big(1 + P_{k}L_{mk}\theta_{k}^{p}R_{k}^{p} \big), \qquad \qquad
\qquad \qquad
\\  m \in \mathcal{U}^p_k \ (k=1, \ldots, K)\end{array}
\end{array} \right.
\end{eqnarray}
where $ \overline{\mathcal {C}}_j^{c} = \log \big (1+\frac{\sum_{k
=1}^{K}P_{k}L_{jk}\theta_{k}^{c}R^{c}}{1+\sum_{k=1}^{K}P_{k}L_{jk}\theta_{k}^{p}R_{k}^{p}}\big),
j\in \mathcal{U}^c$ and $\overline{\mathcal {C}}^{p}_{l}=\log \big( 1 +
\frac{P_{k}L_{lk}\theta_{k}^{c}R^{c} }{1 + P_{k}L_{lk} \theta_{k}^{p}R_{k}^{p}}\big)$ for
$l \in \mathcal{U}^p_k$. $D$ is the total number of streams of OSTBC $\mathbf{S}^c$,
$D_m$ is the number of streams for the $m$-th MS (in the common MS set); $R_k^p$ and
$R^c$ are the encoding rate for the OSTBC $\mathbf S^p_k$ and $\mathbf S^c$ respectively.
\end{Lem}
\proof Please refer to Appendix \ref{pf:lem_SINR} for the proof.
\endproof
\begin{Rem}
The approximation in \eqref{eqn:throughput} is quite tight over a wide range of SNR as
illustrated by Figure \ref{fig:appro}. The physical meaning of $ \overline{\mathcal
{C}}_j^{c}, j\in\mathcal{U}^c$ is the maximum decodable rate for OSTBC $\mathbf{S}^c$  at
the $j$-th MS (in the common MS set) by treating the streams ($\mathbf{S}^p_k,
k=1,\ldots,K$) for the private MSs  as noise and $\overline{\mathcal {C}}^{p}_{l},
l\in\mathcal{U}^p_k$ is the maximum decodable rate at which the $l$th MS (in the $k$-th
private MS set) can successfully decode the OSTBC $\mathbf{S}^c$ by treating the streams
($\mathbf{S}^p_k$) for its own as noise. Hence our detection scheme can always work when
the transmission rate is given in \eqref{eqn:throughput}.
\end{Rem}

\subsection{Long-term Power Allocation Problem Formulation}
It is very important to adjust the long-term power allocation ratio $\{\theta_{k}^{p},
\theta_{k}^{c}\}$ to fully exploit the heterogenous path gain and shadowing effect over
the network MIMO configuration. In this paper, we consider choosing $\{\theta_{k}^{p},
\theta_{k}^{c}\}$ to maximize the {\em minimum weighted throughput (with approximation)}
which is defined as follows.

\begin{Def}[Minimum Weighted Throughput]
\label{def:MW_SINR} Define  $\{w_m, m=1, \ldots, M\}$ to be the positive static
weight\footnote{These QoS weights are determined by the application requirement or the
priority class of the MS and is determined when the communication session is setup.},
which is determined by the Quality-of-Service (QoS) requirement or priority of the $m$-th
MS. The minimum weighted throughput of all the MSs (using the approximation in Lemma
\ref{lem:throughput}) throughput $\mathfrak{C}$ is given by:
\begin{eqnarray}
\mathfrak{C} \big( \{\theta_{k}^{p}, \theta_{k}^{c}\} \big)= \min_{m
\in (\cup_{k=1}^{K}\mathcal{U}^p_k)\cup \mathcal{U}^c}
\{w_m\overline{\mathcal{C}}_m\}
\end{eqnarray}
where $\overline{\mathcal {C}}_{m}$ is given\footnote{In fact, $\overline{\mathcal
{C}}_m$ shall be a function of the power allocation ratio $\{\theta_{k}^{p},
\theta_{k}^{c}\}$. However, we drop them whenever there is no confusion caused through
the rest of the paper for notation convenience.} in \eqref{eqn:throughput}.
\end{Def}

Hence, the optimal long-term power allocation problem can be found
by solving the following optimization problem.
\begin{eqnarray}
\label{eqn:org_prob} \big(\{\theta_{k}^{p,\star},
\theta_{k}^{c,\star}\}\big) = \arg \max_{\{\theta_{k}^{p},
\theta_{k}^{c}\}} & \mathfrak{C} \big( \{\theta_{k}^{p}, \theta_{k}^{c}\} \big) \nonumber \\
\textrm{subject to} & \theta_{k}^{p} + \theta_{k}^{c} = 1, \nonumber \\
& \theta_{k}^{p}, \theta_{k}^{c} \geq 0, \nonumber \\& \forall k \in \{1, \ldots, K\}
\end{eqnarray}

In general, the above optimization problem \eqref{eqn:org_prob} is non-trivial because of
the following reasons. The approximate throughput expression $\overline{\mathcal {C}}_m$
involves complicated operations of the power splitting ratio $\{\theta_{k}^{p},
\theta_{k}^{c}\}$ which is a composition of logarithm and nonlinear SINR expression.
Hence, the objective function is in general non-convex and the standard low complexity
algorithms cannot be directly applied. Moreover, as shown in the current literature, this
type of problems belongs to the minimum throughput maximization problem for the
multi-cell architecture and does not exist a trivial global optimal solution in general
\cite{Tolli}.

\section{Distributive Long-term Power Allocation Algorithms}
\label{sect:sol} In this section, our target is to propose a distributed long-term power
allocation algorithm based on solving the optimization given by \eqref{eqn:org_prob} in a
distributed manner. We formulate the long-term distributive power allocation problem
using a partial cooperative game. We show that the long-term power allocation game has a
unique Nash Equilibrium (NE). Furthermore, we propose a low-complexity distributive
long-term power allocation algorithm which only relies on the local long-term channel
statistics and has provable convergence to the NE.

\subsubsection{Partial Cooperative Game Formulation}

We formulate the long-term power allocation design within the framework of game theory as
a partial-cooperative game, in which the {\em players} are the BSs in the wireless
network and the {\em payoff functions} are the minimum weighted throughput of MSs in the
coverage of each BS. The $k$-th player (BS) competes against the others by choosing his
power allocation ratio $\theta^{c}_k$ given other players' power allocation ratio
\footnote{For fixed $\theta^{c}_k$, $\theta_k^{p}$ can be uniquely determined through
$\theta^{p}_k = 1 - \theta^{c}_k$.} to maximize the minimum weighted throughput
$\mathfrak{C}_k (\theta^{c}_k, \boldsymbol{\theta^{c}_{-k}})$, which is given by
\begin{eqnarray}
\mathfrak{C}_k (\theta^{c}_k, \boldsymbol{\theta}_{-k}^{c}) = \min
\Big\{ f_{k}^{1}(\theta_{k}^{c},\boldsymbol{\theta}_{-k}^{c}),
f_{k}^{2}(\theta_{k}^{c},\boldsymbol{\theta}_{-k}^{c}) \Big\}
\label{eqn:loc_uti}
\end{eqnarray}
where
\begin{eqnarray}
f_{k}^{1}(\theta_{k}^{c},\boldsymbol{\theta}_{-k}^{c})  & = & \min
\Big\{\min_{j \in \mathcal{U}^c}\{w_j
\frac{D_j}{D}\overline{\mathcal C}^c_{j}\}, \min_{j \in
\mathcal{U}^c}\{w_j \frac{D_j}{D} \overline{\mathcal C}^p_l,
l\in\mathcal{U}^p_k\} \Big\} \notag\\
& = & \min \Big\{\underbrace{\min_{j \in
\mathcal{U}^c}\{w_j\frac{D_j}{D}\overline{\mathcal
C}^c_{j}\}}_{g^{1}(\boldsymbol{\theta}^{c})}, \underbrace{\min_{j
\in \mathcal{U}^c}\{w_j\frac{D_j}{D}\}\overline{\mathcal C}^p_l,
l\in\mathcal{U}^p_k}_{g_{k}^{2}({\theta}_{k}^{c})} \Big\}
\end{eqnarray} is the minimum weighted throughput (with
approximation) of all the MSs in the common MS set,
\begin{eqnarray}
f_{k}^{2}(\theta_{k}^{c},\boldsymbol{\theta}_{-k}^{c}) & = & w_m
\overline{\mathcal C}_{m}, m\in \mathcal{U}^p_k \notag \\
& = & {w_{m}}\log\left ( 1+P_{k}L_{mk}(1-\theta_{k}^{c})R_{k}^{p}
\right ), m\in \mathcal{U}^p_k
\end{eqnarray}
is the weighted throughput (with approximation) of the MS in the $k$-th private MS set
and $\boldsymbol{\theta}^{c}_{-k} \triangleq (\theta_q^{c})_{q = 1, \ q \neq k}^{K}$ is
the set of long-term power allocation ratios of all the BSs except the $k$-th one. Hence,
the partial-cooperative game is formulated as:
\begin{eqnarray} \label{game}
(\mathscr{G}) : \quad \left.
\begin{array}{c c}
\max_{\theta_k^{c}} & \mathfrak{C}_k (\theta^{c}_k, \boldsymbol{\theta}^{c}_{-k}) \\
\textrm{subject to} & \boldsymbol{\theta}_k \in \mathscr{D}_k
\end{array}
\right. \quad \forall k \in \mathscr{K},
\end{eqnarray}
where $\mathscr{K} \triangleq \{1, \ldots, K\}$ denotes the set of
all players, i.e. the BSs, $\mathfrak{C}_k (\theta^{c}_k,
\boldsymbol{\theta}^{c}_{-k})$ defined in \eqref{eqn:loc_uti} is the
payoff functions of the player $k$ and $\mathscr{D}_k$ is the
admissible strategy set for player $k$, defined as $\mathscr{D}_k
\triangleq \{\theta \in \mathbb R : 0 \leq \theta \leq 1\}$. The
solutions of the above game $\mathscr{G}$ are formally defined as
follows.

\begin{Def}[Nash Equilibrium]
A (pure) strategy profile of the long-term power allocation
$\boldsymbol{\theta}^{c,\star} = (\theta_k^{c,\star})_{k \in
\mathscr{K}} \in \mathscr{D}_k \times \ldots \times \mathscr{D}_K$
is a NE of the game $\mathscr{G}$ if
\begin{eqnarray}
\mathfrak{C}_k (\theta^{c, \star}_k,
\boldsymbol{\theta}^{c,\star}_{-k}) \geq \mathfrak{C}_k
(\theta^{c}_k, \boldsymbol{\theta}^{c,\star}_{-k}), \quad \forall
\theta_k^{c} \in \mathscr{D}_k, \ \forall k \in \mathscr{K}.
\label{eqn:NE}
\end{eqnarray}
\end{Def}

At a NE point, each BS $k$, given the long-term power allocation
profile of other BSs $\boldsymbol{\theta}_{-k}^{c,\star}$, cannot
improve its utility (or payoff) by unilaterally changing his own
long-term power allocation strategy $\theta_{k}^{c}$. The absence of
NE simply means that the distributed system is inherently unstable.
In order to obtain the distributed solution, we shall characterize
the properties of NE such as the existence and uniqueness through
the following theorem.

\begin{Thm}[Existence and Uniqueness of NE]
\label{thm:nash_exist} The strategic partial-cooperative game
$\mathscr{G}$ has the following two properties:
\begin{enumerate}
\item There exists a NE for the strategic partial-cooperative game $\mathscr{G}$.
\item Moreover, the NE is unique.
\end{enumerate}
\end{Thm}
\proof Please refer to Appendix \ref{pf:thm_nash_exist} for the proof. \endproof

From Theorem \ref{thm:nash_exist}, we can well establish the
properties of the strategic partial-cooperative game $\mathscr{G}$,
which is shown to have a unique NE, and hence the distributed system
is stable. In the following part, we shall develop the corresponding
distributive algorithm to achieve the optimal long-term power
allocation for each BS in the overlaid network architecture.

\subsubsection{Algorithm Description}
Since the above partial-cooperative game $\mathscr{G}$ has a unique
NE point, we shall try to develop the proper algorithm to find the
desired point. Traditionally, in a partial-cooperative game, the
best-response update algorithm \cite{Fudenberg91} is shown to
achieve good performance. However, the convergence property cannot
be guaranteed due to the reason that there might be some best
response cycles \cite{Voorneveld}. In our partial-cooperative game,
the best response update can be shown to be not converging to the NE
in some cases. In what follows, we shall propose a novel algorithm
with provable convergence property.

\begin{algorithm}
\caption{Distributive Long-term Power Allocation Algorithm}
\label{alg2:Game}
\begin{itemize}
\item \textbf{Step 1: Initialization}:

Set iteration index $i = 1$.

For each base station $\mathit{l}\in\mathscr{L}$ choose some power
the splitting ratio $\eta_{l}^{c,*} \in [0,1)$ and $\eta_{l}^{p,*} =
1 - \eta_{l}^{c,*}$ such that
$g_{l}^{2}\big(\eta_{l}^{c,*}\big)=f_{l}^{2}\big(\eta_{l}^{c,*}\big)$.
Under this initial power allocation setup $\boldsymbol{\eta^{c,*}} =
[\eta_{1}^{c,*}, \eta_{2}^{c,*}, \ldots, \eta_{L}^{c,*}]$, the
common MS broadcasts the measured receive SINR $\Gamma_{\min} (1)$
to all the cooperative BSs.

Each BS solve the equation
$w^{c}\log(1+\Gamma_{\min}(1))=w^{p}_{l}\log(1+P_{l}L_{l}^{p}(1-\theta_{l}^{c})R_{l}^{p})$
with respect to $\theta_{l}^{c}$ to get the root $\xi_{l}^{c}(1)$.
Each BS adjust its own power allocation ratio such that
$\theta_{l}^{c} (1) =\max\{\xi_{l}^{c}(1),\eta_{l}^{c,*}\}$. Under
the power allocation vector $\boldsymbol{\theta^{c}}(1) =
[\theta_{1}^{c}(1), \theta_{2}^{c}(1), \ldots, \theta_{L}^{c}(1)]$,
the common MS broadcasts the measured receive SINR
$\Gamma_{\max}(1)$ to all the BSs.

\item \textbf{Step 2: Power Allocation Update}:

Update index $i := i+1$

We choose
$\Gamma(i)=\big(\Gamma_{\min}(i-1)+\Gamma_{\max}(i-1)\big)/2$, each
BS solves the equation
$w^{c}\log(1+\Gamma(i))=w^{p}_{l}\log(1+P_{l}L_{l}^{p}(1-\theta_{l}^{c})R_{l}^{p})$
to get the solution $\xi_{l}^{c}(i)$. Each BS adjusts its own power
allocation ratio
$\theta_{l}^{c}(i)=\max\{\xi_{l}^{c}(i),\eta_{l}^{c,*}\}$.

\item \textbf{Step 3: Signal and Interference Plus Noise
Estimation}:

The common MS broadcasts the measured receive SINR
$\overline{\Gamma(i)}$ to all the base stations.

\item \textbf{Step 4: Termination}:

If $\Gamma(i) < \overline{\Gamma(i)}$,
$\Gamma_{\min}(i)=\Gamma(i)$ and $\Gamma_{\max}(i)=\Gamma_{\max}(i-1)$.

If $\Gamma(i) > \overline{\Gamma(i)}$, $\Gamma_{\max}(i) =\Gamma(i)$ and $\Gamma_{\min}(i)=\Gamma_{\min}(i-1)$.

If $\Gamma(i)=\overline{\Gamma(i)}$, terminate; otherwise, go to Step 2.

\end{itemize}
\end{algorithm}

\begin{Thm} \label{thm:algorithm_convergence}
The proposed iterative power allocation algorithm in Algorithm 2
converges to the unique NE defined by \eqref{eqn:NE}.
\end{Thm}
\proof Please refer to Appendix \ref{pf:thm_algorithm_convergence} for the proof. \endproof

\begin{figure}
\centering
\includegraphics[width = 5in]{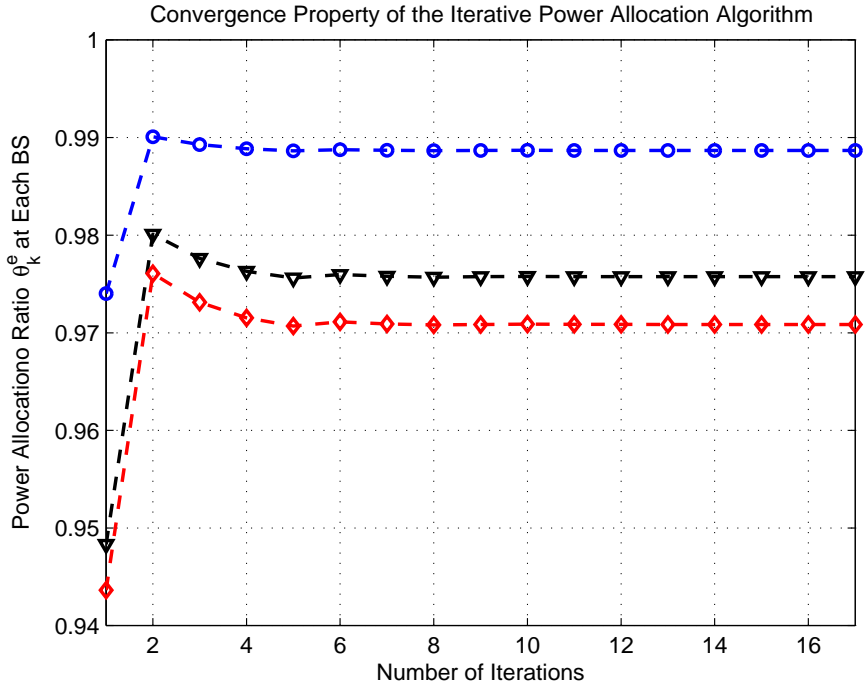}
\caption{Convergence property of the proposed distributive long-term power allocation
algorithm in a cellular network with 3 BSs. Assume the private MS sets and the common MS
set have already been determined with our user scheduling algorithm. As a result, 3 BSs
non-cooperatively serve 3 private MSs (MS1,MS2 and MS3) respectively and two ($M^c=2$)
common MS (MS4 and MS5) cooperatively. In the simulation, we choose
$P_{1}=P_{2}=P_{3}=30dBm; L_{11} = -118.30dB, L_{12} = -140.14dB, L_{13} = -139.29dB;
L_{21} = -145.11dB, L_{22} = -115.56dB, L_{23} = -143.23dB; L_{31} = -147.78dB, L_{32} =
-139.65dB, L_{33} = -116.35dB; L_{41} = -135.24dB, L_{42} = -136.08dB, L_{43} =
-135.35dB; L_{51} = -135.16dB, L_{52} = -135.91dB, L_{53} = -134.94dB$ and the QoS
weighting coefficients are given by $w_{1} = w_{5} = 2, w_{2}=w_{3}=w_{4}=1$.}
\label{fig:convergence}
\end{figure}

\begin{Rem}[Complexity and Signaling Overhead] The computation of
the {\em long-term power allocation} is distributed at each of the
$K$ BSs. Furthermore, the iterations are done over a long time scale
(instead of short-term CSI time scale) where each common MS shall
broadcast its long-term SINR, which is a scalar, in each iteration.
As a result, the signaling overhead is very low. Furthermore, as
illustrated in Fig. \ref{fig:convergence}, the algorithm has fast
convergence and hence, the total iterations required are very
limited.
\end{Rem}

Fig. \ref{fig:convergence} shows the convergence property of the proposed power
allocation algorithm as specified in the figure caption. Through the numerical studies,
we found that the proposed long-term power allocation algorithm is shown to have a fast
convergence.

\section{Numerical Examples}
\label{sect:experiments} This section provides some numerical examples to verify the
behavior of our proposed long-term power allocation strategies in a network MIMO
configuration with $10$ cells arranged in a hexagonal manner. Each cell has 5km radius
and there are 50 type-I MSs (regular MS with weight 1) and 50 type-II MSs (higher
priority MSs with weight 2) in the system. These MSs are moving with high speed in the
coverage according to a random-walk model\cite{Mobility_Model}. We also assume each BS is
equipped with $4$ antennas, namely $N_t = 4$, and $N_t^{c} = N_t^{e} = 2$. Each MS is
equipped with $2$ antennas, namely $N_r=2$. The path gain model is given by
$\textrm{PG(dB)} = -130.19 - 37.6\log_{10}(d\textrm{(km)})$ and shadowing standard
deviation is $8$dB as specified in the IEEE 802.16m evaluation
methodology\cite{Std:4G-EMD:16m}. For illustration purpose, we compare our proposed
long-term power allocation algorithm with the following baseline schemes. {\em Baseline
1: Orthogonal-Division (TDD/FDD) Based Strategy}, i.e., the BSs serve the private and
common MSs alternatively in different time/frequency slots. {\em Baseline 2: Uniform
Power Allocation}, i.e., the BSs serve the private and common MSs simultaneously with
uniform power allocation. {\em Baseline 3: Centralized Long-term Power Allocation}, i.e.,
the BSs serve the private and common MSs simultaneously using our open-loop SDMA scheme
but the long-term power allocation is computed by the brute-force centralized numerical
evaluations of problem \eqref{eqn:org_prob}.

\begin{figure}
\centering
\includegraphics[width= 5in]{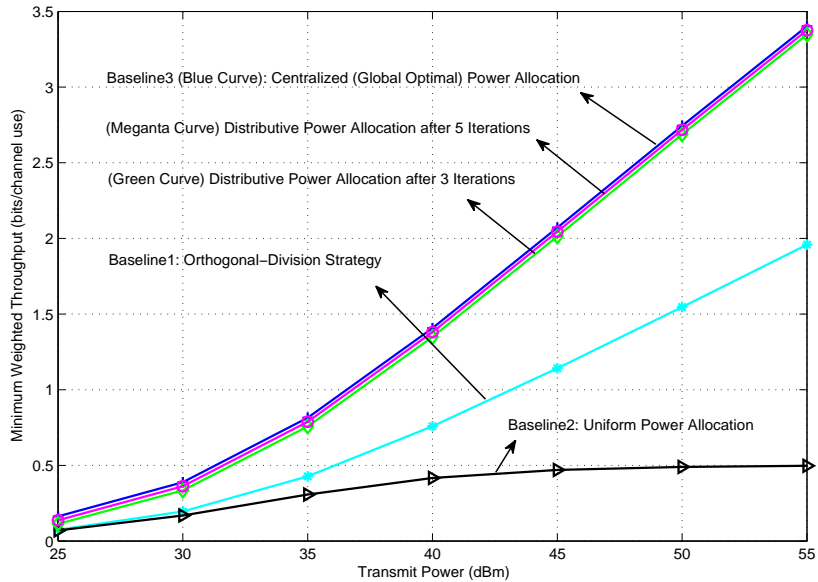}
\caption{Minimum weighted throughput comparison for different open-loop schemes with
respect to the transmit power. Through the numerical examples, the minimum weighted
throughput of the proposed distributive long-term power allocation scheme with finite
iteration numbers can outperform the conventional orthogonal-division (TDD/FDD) based
open-loop scheme (baseline 1) as well as the open-loop overlaying scheme with uniform
power allocation (baseline 2) and has negligible performance loss compared to the
open-loop overlaying scheme with centralized (global optimal) power allocation (baseline
3). In the simulation, we choose $\xi^p_1 = \ldots = \xi^p_K = 20$dB and $\xi^c =5$dB.}
\label{fig:throughput_power}
\end{figure}

Fig. \ref{fig:throughput_power} shows the minimum weighted throughput comparison for
different open-loop schemes with
respect to the BS transmit power. From the numerical examples, we notice that if we apply
the open-loop overlaying scheme in a naive manner without careful long-term power
allocation, the system performance can even be worse than the traditional
orthogonal-division based open-loop scheme (baseline 1 over 2). However, if we utilize
the open-loop overlaying scheme with careful long-term power control, the system
performance can be greatly improved (baseline 3 and the proposed scheme over baseline 1
and 2). Moreover, the proposed low complexity distributive power allocation algorithm can
achieve significant performance advantage over the traditional orthogonal-division
(TDD/FDD) schemes (baseline 1) and has negligible performance loss with respect to the
centralized scheme (baseline 3). From the simulation results, we observe that the
performance of the proposed distributive long-term power allocation algorithm is
close-to-optimal (baseline 3). In other words, the NE of the partial cooperative game in
\eqref{game} is quite "social optimal\footnote{The near social optimal performance of the
NE is due to the {\em partial cooperative component} in the game formulation in
\eqref{game}.}".
%\textcolor{red}{Since the path loss from the BSs to the common MS may not be the same, it may help to further improve performance if we allow the common MS to select a subset of serving BSs. To illustrate the potential benefit, we have simulated the performance of BS selection for the common MS using exhaustive search in Fig. \ref{fig:throughput_power}. It is shown that only marginal gain can be achieved compared with the proposed design. This is because by Algorithm 1, the path loss between the BSs and the common MS will not be too asymmetric.}

\begin{figure}
\centering
\includegraphics[width=5in]{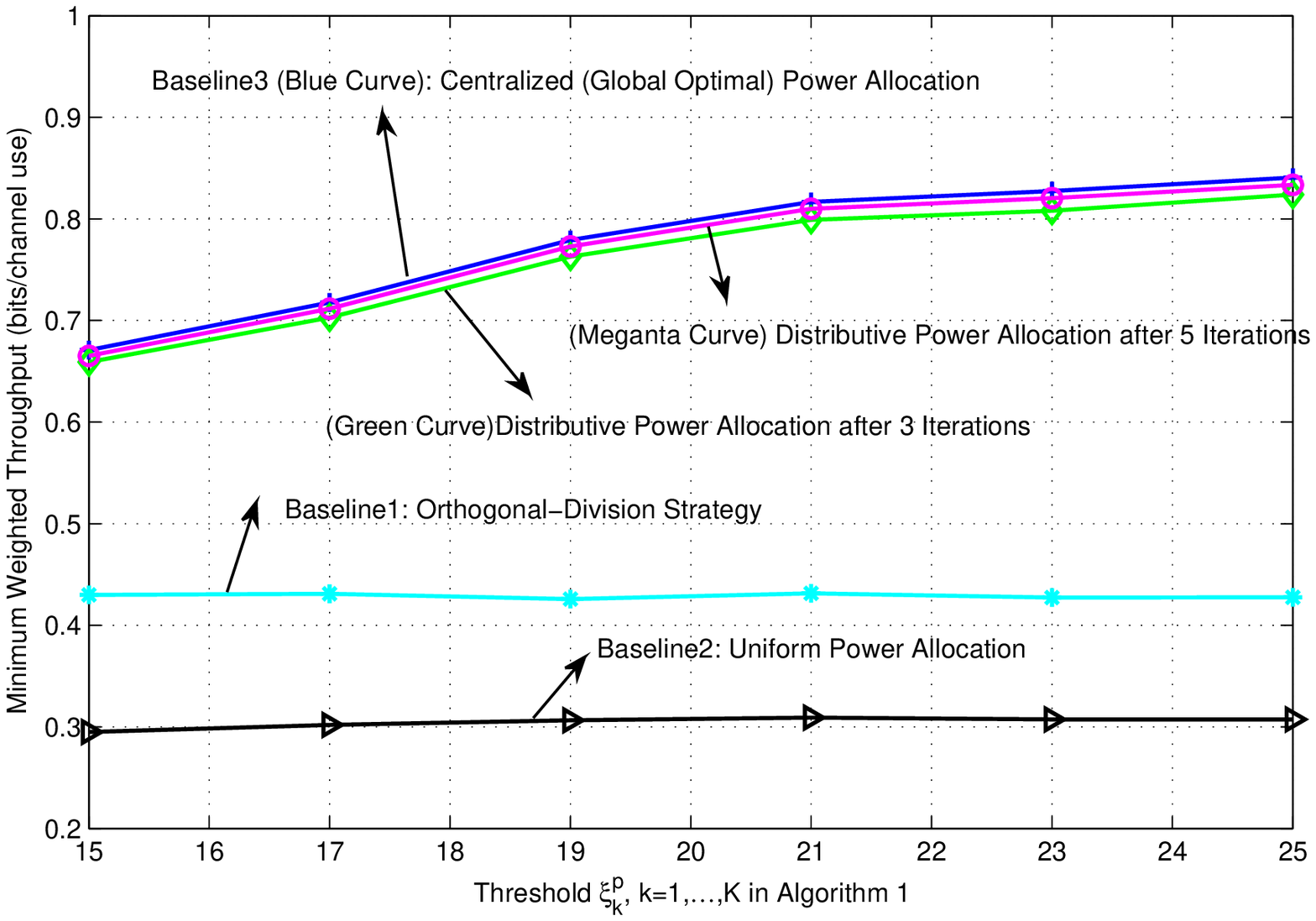}
\caption{Minimum weighted throughput comparison for different open-loop schemes with
respect to the private MS set threshold $\xi^p_k$ in Algorithm 1. Through the numerical
examples, the minimum weighted throughput of the proposed distributive long-term power
allocation scheme with finite iteration numbers can outperform the conventional
orthogonal-division (TDD/FDD) based open-loop scheme (baseline 1) as well as the
open-loop overlaying scheme with uniform power allocation (baseline 2) and has negligible
performance loss compared to the open-loop overlaying scheme with centralized (global
optimal) power allocation (baseline 3). In the simulation, we keep $\xi^p_1 = \ldots =
\xi^p_K$ and choose $\xi^c =5$dB. Transmit Power of all the BSs are 35dBm. }
\label{fig:throughput_private_threshold}
\end{figure}

\begin{figure}
\centering
\includegraphics[width=5in]{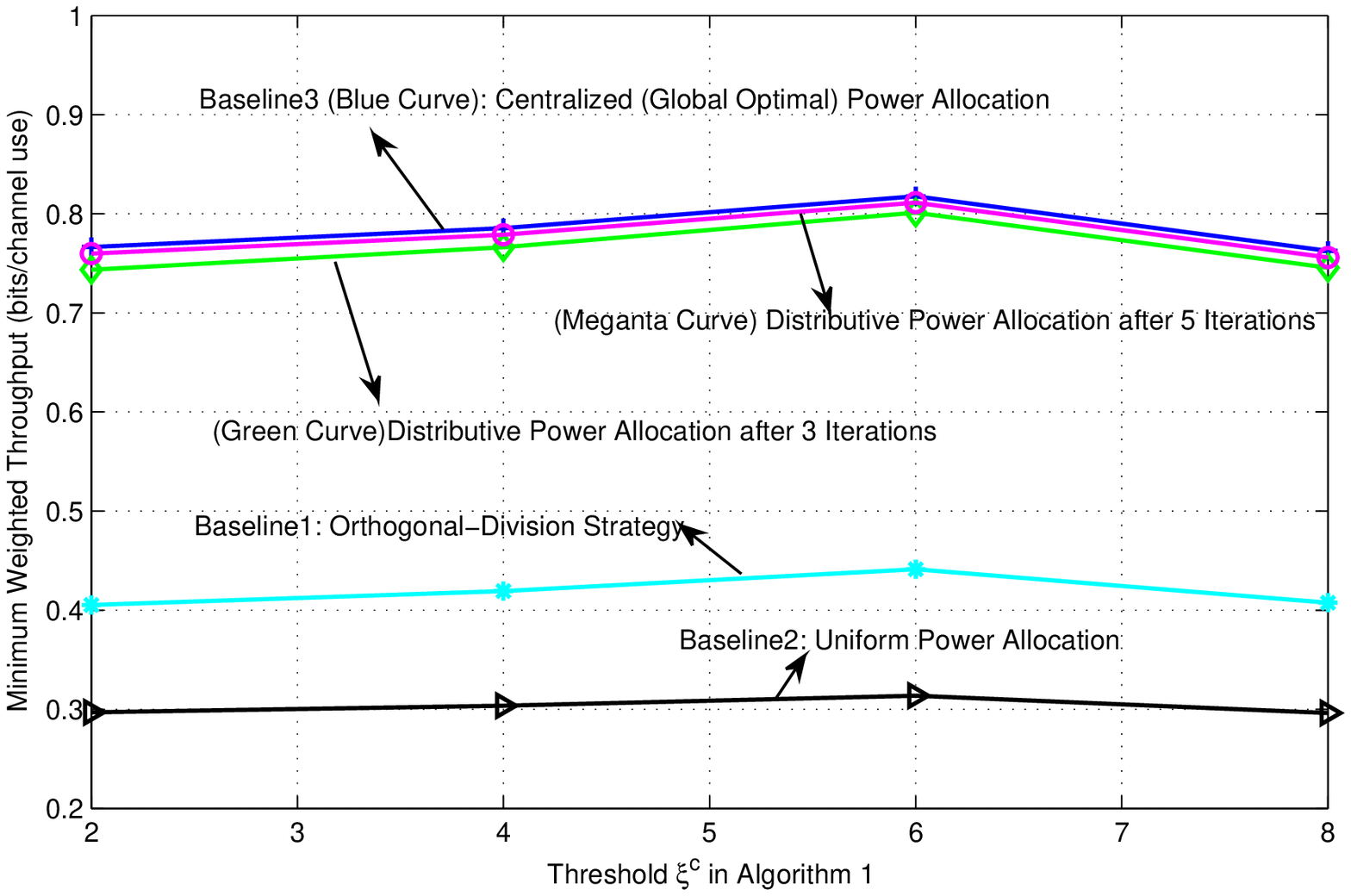}
\caption{Minimum weighted throughput comparison for different open-loop schemes with
respect to the common MS set threshold $\xi^c$ in Algorithm 1. Through the numerical
examples, the minimum weighted throughput of the proposed distributive long-term power
allocation scheme with finite iteration numbers can outperform the conventional
orthogonal-division (TDD/FDD) based open-loop scheme (baseline 1) as well as the
open-loop overlaying scheme with uniform power allocation (baseline 2) and has negligible
performance loss compared to the open-loop overlaying scheme with centralized (global
optimal) power allocation (baseline 3). In the simulation, we choose $\xi^p_1 = \ldots =
\xi^p_K = 20dB$. Transmit Power of all the BSs are 35dBm. }
\label{fig:throughput_common_threshold}
\end{figure}

Fig. \ref{fig:throughput_private_threshold} and Fig.
\ref{fig:throughput_common_threshold} show the minimum weighted throughput comparison for
different open-loop schemes with respect to the private and common MS set thresholds
$\xi^p_k$ and $\xi^c$ in Algorithm 1 respectively. We observe
the performance gain of the proposed open-loop overlaying scheme over various baselines
at various threshold values.

Moreover, the numerical result in Fig. \ref{fig:throughput_private_threshold} and Fig.
\ref{fig:throughput_common_threshold} again show that the NE of our partial cooperative
game in \eqref{game} is almost "social optimal".

\section{Conclusion}
\label{sect:conclusion} In this paper, we proposed an open-loop overlaying scheme to
adjust the dynamic and heterogeneous MIMO configurations in the network MIMO systems. To
exploit the heterogeneous path gain effect among multiple cells, we propose a
distributive low complexity long-term power allocation algorithm with provable
convergence property which only relies on local channel statistics. Through numerical
studies, we show that the proposed open-loop overlaying scheme with distributive
long-term power allocation algorithm can achieve significant performance advantages over
the traditional schemes and has negligible performance loss compared with the centralized
scheme.

\appendices
\section{Proof of Lemma \ref{lem:throughput}}
\label{pf:lem_SINR}Denote $\mathcal C^{c}$ to be transmission data rate of the OSTBC
$\mathbf S^c$, also denote $\mathcal {C}^{p}_{l}, l\in \mathcal{U}^p_k$ to be the maximum
decodable rate at which the $l$-th private MS ($l\in\mathcal{U}^p_k$) can successfully
decode the steams for the common MS set, i.e. OSTBC $\mathbf S^c$. The expression of
$\mathcal {C}{p}_{l}$ is given by:
\begin{eqnarray}
\mathcal {C}^{p}_{l} = \mathbb{E} \big[\log (1 + \gamma^{p}_l) \big]
\end{eqnarray}
where $\gamma_l^p$ is the instantaneous SINR at the $l$-th MS (in
the $k$-th private MS set) when decoding OSTBC $\mathbf S^c$ by
treating OSTBC $\mathbf S_k^p$ as noise and is given by:
\begin{eqnarray}
\gamma_l^p = \frac{\textrm{Tr} \left\{ \left(
\sqrt{\frac{P_{k}L_{lk}\theta_{k}^{c}}{N_{t}^{c}}}\mathbf{H}_{lk,2}
 \mathbf{S}^{c} \right) {\left(
\sqrt{\frac{P_{k}L_{lk} \theta_{k}^{c}}{N_{t}^{c}}}\mathbf{H}_{lk,2}
\mathbf{S}^{c} \right)}^{H} \right \}}{\textrm{Tr} \left\{\left(
\sqrt{\frac{P_{k}L_{lk}\theta_{k}^{p}}{N_{t}^{p}}} \mathbf H_{lk,1}
\mathbf S_{k}^{p} + \mathbf Z_l \right){\left(
\sqrt{\frac{P_{k}L_{lk}\theta_{k}^{p}}{N_{t}^{p}}} \mathbf H_{lk,1}
\mathbf S_{k}^{p} + \mathbf Z_l \right)}^{H} \right\}} \qquad l\in
\mathcal{U}^p_k
\end{eqnarray}
where $\textrm{Tr}(\cdot)$ denotes the matrix trace operation. Hence
SIC can be performed at the $l$-th MS only when $\mathcal C^{c} \leq
\mathcal {C}^p_{l}$ and the achievable throughput for the $l$-th MS
after SIC is given by
\begin{eqnarray}
\mathcal {C}_{l} = \mathbb{E} \big[\log (1 + \lambda_l) \big] \qquad
l\in\mathcal{U}^p_k
\end{eqnarray}
where $\lambda_l$ denotes the instantaneous SNR (after the
interference cancelation) at $l$-th MS (in the $k$-th private MS
set) and is given by:
\begin{eqnarray}
\lambda_l  =   \frac{ \textrm{Tr} \big\{ (
\sqrt{\frac{P_{k}L_{lk}\theta_{k}^{p}}{N_{t}^{p}}} \mathbf
H_{lk,1}\mathbf S_{k}^{p} )(
\sqrt{\frac{P_{k}L_{lk}\theta_{k}^{p}}{N_{t}^{p}}} \mathbf
H_{lk,1}\mathbf S_{k}^{p} )^{H}\big\}}{\textrm{Tr} \big\{ \mathbf
Z_{l} \mathbf Z_{l}^{H}\big\}}, \qquad l\in \mathbf{U}^p_k
\end{eqnarray}

Let $\mathcal C_j^{c} = \mathbb{E} \big[\log (1 + \gamma^c_j) \big], j\in\mathcal{U}^c$
be the maximum decodable rate for the OSTBC $\mathbf S^c$ at the $j$-th MS (in the common
MS set) with $\gamma^c_j$ denoting the instantaneous SINR at the $j$-th MS (in the common
MS set). The expression of $\gamma^c_j$ is given by:
\begin{eqnarray}
\gamma^c_j & = & \frac{\textrm{Tr} \left\{ \left(
\sum_{k=1}^{K}\sqrt{\frac{P_{k}L_{jk}\theta_{k}^{c}}{N_{t}^{c}}}\mathbf{H}_{jk,2}
\mathbf{S}^{c} \right) {\left(
\sum_{k=1}^{K}\sqrt{\frac{P_{k}L_{jk}\theta_{k}^{c}}{N_{t}^{c}}}\mathbf{H}_{jk,2}
\mathbf{S}^{c} \right)}^{H} \right \}}{\textrm{Tr} \left\{\left(
\sum_{k=1}^{K} \sqrt{\frac{P_{k}L_{jk}\theta_{k}^{p}}{N_{t}^{p}}}
\mathbf H_{jk,1} \mathbf S_{k}^{p} + \mathbf Z_j \right){\left(
\sum_{k=1}^{K} \sqrt{\frac{P_{k}L_{jk}\theta_{k}^{p}}{N_{t}^{p}}}
\mathbf H_{jk,1} \mathbf S_{k}^{p} + \mathbf Z_j\right)}^{H}
\right\}}
\end{eqnarray}
If we choose the transmission rate of OSTBC $\mathbf S^c$ to be $\mathcal C^{c} =
\min_{j\in \mathcal{U}^c, l\in\mathcal{U}^p_k, k=1,\ldots,K} \{\mathcal C_j^{c}, \mathcal
C^{p}_l\}$, all the private MSs can successfully perform SIC and the result throughput
expression for the MSs in the common MS set and private MS sets is given by
\begin{eqnarray}
\mathcal {C}_m = \left\{ \begin{array}{ll} \frac{D_m}{D} \mathcal
C^{c} = \frac{D_m}{D} \min_{j\in \mathcal{U}^c, l\in\mathcal{U}^p_k,
k=1,\ldots,K} \{\mathcal C_j^{c}, \mathcal C^{p}_l\}, &\qquad m\in \mathcal{U}^c \\
\mathbb{E} \big[\log (1 + \lambda_m) \big], &\qquad m\in
\mathcal{U}^p_k, k=1,\ldots,K \end{array} \right.
\end{eqnarray}
where $\mathcal C_j^{c}, j\in\mathcal{U}^c$; $\mathcal C^{p}_l, l\in \mathcal{U}^p_k,
k=1, \ldots, K$ and $\mathbb{E} \big[\log (1 + \lambda_m) \big], m\in \mathcal{U}^p_k,
k=1,\ldots,K$ are given by:
\begin{eqnarray}
\mathcal {C}_j^{c} & = & \mathbb{E}  \left [\log \big(
1+\gamma_j^{c}\big) \right]  \\
& \approx & \log \bigg(1+\frac{\mathbb E \left[ \textrm{Tr} \left\{
\left(\sum_{k=1}^{K}\sqrt{\frac{P_{k}L_{jk}\theta_{k}^{c}}{N_{t}^{c}}}\mathbf{H}_{jk,2}
\mathbf{S}^{c} \right) {\left(
\sum_{k=1}^{K}\sqrt{\frac{P_{k}L_{jk}\theta_{k}^{c}}{N_{t}^{c}}}\mathbf{H}_{jk,2}
\mathbf{S}^{c} \right)}^{H} \right\} \right]}{\mathbb E \left[
\textrm{Tr} \big\{ \big( \sum_{k=1}^{K}
\sqrt{\frac{P_{k}L_{jk}\theta_{k}^{p}}{N_{t}^{p}}} \mathbf H_{jk,1}
\mathbf{S}_{k}^{p} + \mathbf Z_j \big){\big( \sum_{k=1}^{K}
\sqrt{\frac{P_{k}L_{jk}\theta_{k}^{p}}{N_{t}^{p}}} \mathbf H_{jk,1}
\mathbf S_{k}^{p} + \mathbf Z_j\big)}^{H} \big\} \right]} \bigg)\\
& = & \log \bigg(1+\frac{\sum_{k=1}^{K} \mathbb{E} \left[
\textrm{Tr} \left\{ \frac{P_{k}L_{jk}\theta_{k}^{c}}{N_{t}^{c}}
\mathbf{H}_{jk,2}\mathbf{S}^{c} \mathbf{S}^{c,H}
\mathbf{H}_{jk,2}^{H} \right\} \right ]}
{\mathbb{E}\left[\textrm{Tr}
\left(\mathbf{Z}_j\mathbf{Z}_j^{H}\right) \right] +
\sum_{k=1}^{K}\mathbb{E}\left[
\textrm{Tr}\left(\frac{P_{k}L_{jk}\theta_{k}^{p}}{N_{t}^{p}}\mathbf{H}_{jk,1}\mathbf{S}_{k}^{p}\mathbf{S}_{k}^{p,H}\mathbf{H}_{jk,1}^{H}\right)
\right]}\bigg), \quad j\in \mathcal{U}^c \label{eqn:gamma_0}
\end{eqnarray}

\begin{eqnarray}
\mathcal {C}^{p}_{l} & = & \mathbb{E}  \left [\log \big( 1 + \gamma^{p}_l\big) \right] \label{eqn:appro0}\\
& \approx & \log \bigg(1+\frac{\mathbb E \left[ \textrm{Tr} \left\{ \left(
\sqrt{\frac{P_{k}L_{lk}\theta_{k}^{c}}{N_{t}^{c}}}\mathbf{H}_{lk,2}
 \mathbf{S}^{c} \right) {\left(
\sqrt{\frac{P_{k}L_{lk}\theta_{k}^{c}}{N_{t}^{c}}}\mathbf{H}_{lk,2} \mathbf{S}^{c}
\right)}^{H} \right\} \right]}{\mathbb E \left[ \textrm{Tr} \left\{\left(
\sqrt{\frac{P_{k}L_{lk}\theta_{k}^{p}}{N_{t}^{p}}} \mathbf H_{lk,1} \mathbf S_{k}^{p} +
\mathbf Z_{l} \right){\left( \sqrt{\frac{P_{k}L_{lk}\theta_{k}^{p}}{N_{t}^{p}}} \mathbf
H_{lk,1}\mathbf S_{k}^{p} + \mathbf Z_{l}\right)}^{H} \right\} \right]} \bigg) \label{eqn:appro}\\
& = & \log \bigg(1+\frac{ \mathbb{E} \left[ \textrm{Tr} \left\{
\frac{P_{k}L_{lk}\theta_{k}^{c}}{N_{t}^{c}} \mathbf{H}_{lk,2}\mathbf{S}^{c}
\mathbf{S}^{c,H} \mathbf{H}_{lk,2}^{H} \right\} \right ]}{\mathbb{E}\left[\textrm{Tr}
\left(\mathbf{Z}_l\mathbf{Z}_l^{H}\right) \right]+\mathbb{E}\left[
\textrm{Tr}\left(\frac{P_{k}L_{lk}\theta_{k}^{p}}{N_{t}^{p}}\mathbf{H}_{lk,1}\mathbf{S}_{k}^{p}\mathbf{S}_{k}^{p,H}\mathbf{H}_{lk,1}^{H}\right)
\right]}\bigg), \quad l\in \mathcal{U}_k^c, k=1, \ldots, K\label{eqn:gamma_2}
\end{eqnarray}

\begin{eqnarray}
\mathbb{E}  \left [\log \big( 1+\lambda_m\big) \right] & \approx &
\log \bigg( 1+\frac{\mathbb E \big[ \textrm{Tr}
(\frac{P_{k}L_{mk}\theta_{k}^{p}}{N_{t}^{p}} \mathbf H_{mk,1}
\mathbf S_{k}^{p} \mathbf S_{k}^{p,H} \mathbf H_{mk,1}^{H})
\big]}{\mathbb E \big[\textrm{Tr}(\mathbf Z_{m} \mathbf
Z_{m}^{H})\big] } \bigg ), m \in \mathcal{U}^p_k, k=1, \ldots,
K\label{eqn:gamma_l}
\end{eqnarray}
where  the approximation is asymptotically tight when the matrix
size $N_t^{p}$ is sufficiently large. Fig. \ref{fig:appro}
illustrates the quality of the approximation of \eqref{eqn:appro}
for reasonable values of $N_t^{p}$ (4x4 MIMO configurations). As
shown in the figure, the quality of the approximation is quite good
even for small $N_t^{p}$. Equations \eqref{eqn:gamma_0},
\eqref{eqn:gamma_l} and \eqref{eqn:gamma_2} are derived from the
independent property between the noise and the transmitted signals.

Due to the OSTBC transmission structure at the BS side, the above relations can be
simplified and we have \cite{Tarokh,Tarokh_diversity}
\begin{eqnarray}
\mathbf S_k^{p} \mathbf S_k^{p,H} & = & \sum_{j=1}^{R^{p}_k T}
\|s_{k,j}^{p}\|^{2}\mathbf{I}_{N_t^p} = R_k^p T \mathbf{I}_{N_t^p}, \forall k = 1, 2, \ldots, K\\
\mathbf S^{c} \mathbf S^{c,H} & = & \sum_{j=1}^{R^{c}T}
\|s_{j}^{c}\|^{2}\mathbf{I}_{N_t^c} = R^{c} T \mathbf{I}_{N_t^c}
\end{eqnarray}
where we use the assumption that the transmitted symbols are
normalized to unity, i.e., $ \|s_{k,j}^{p}\|^{2} = \|s_{j}^{p}\|^{2}
= 1$ for all $k$. Hence, the following relations can be directly
derived.
\begin{eqnarray}
\mathbb{E} \left[ \textrm{Tr} \left( \frac{P_{k}L_{mk}\theta_{k}^{c}}{N_{t}^{c}} \mathbf{H}_{mk,2}\mathbf{S}^{c} \mathbf{S}^{c,H} \mathbf{H}_{mk,2}^{H} \right) \right ] & = & P_{k}L_{mk}\theta_{k}^{c}R^{c}TN_{r} \label{eqn:center_signal}\\
\mathbb{E}\left[
\textrm{Tr}\left(\frac{P_{k}L_{mk}\theta_{k}^{p}}{N_{t}^{p}}\mathbf{H}_{mk,1}\mathbf{S}_{k}^{p}\mathbf{S}_{k}^{p,H}\mathbf{H}_{mk,1}^{H}\right)
\right] & = & P_{k}L_{mk}^{c}\theta_{k}^{p}R_{k}^{p}TN_{r}\\
\mathbb E \big[\textrm{Tr}(\mathbf Z_m^{c} \mathbf Z_m^{c,H}) \big] & = & N_{r}T
\end{eqnarray}
Substitute the above relations into \eqref{eqn:gamma_0}, \eqref{eqn:gamma_2} and
\eqref{eqn:gamma_l} and denote the approximate value of $\mathcal C_m, \mathcal{C}_j^{c},
\mathcal {C}^{p}_{l}$ and $\mathbb{E} \left [\log \big( 1+\lambda_m\big) \right]$ with
$\overline{\mathcal C}_m, \overline{\mathcal {C}}_j^{c}, \overline{\mathcal {C}}^{p}_{l}$
and $\log \big(1 + P_{k}L_{mk}\theta_{k}^{p}R_{k}^{p} \big)$ respectively, we have Lemma
\ref{lem:throughput}.

\section{Proof of Theorem \ref{thm:nash_exist}}
\label{pf:thm_nash_exist} We shall provide the proofs of the existence and uniqueness in the
following two subsections.
\subsection{Existence}
To prove the existence of the NE in the game $\mathscr{G}$, we shall first characterize the
following two properties of the payoff functions in the game $\mathscr{G}$.
\begin{enumerate}
\item{The payoff function $\mathfrak{C}_{k} (\theta^{c}_k, \boldsymbol{\theta}_{-k}^{c})$ is continuous in
$\boldsymbol{\theta}^{c}$.}

We first notice that $w_j\overline{\mathcal C}^{c}_{j}, j\in
\mathcal{U}^c$,
$g_{k}^{2}(\theta_{k}^{c},\boldsymbol{\theta}_{-k}^{c})$ and
$f_{k}^{2}(\theta_{k}^{c},\boldsymbol{\theta}_{-k}^{c})$ are
continuous functions with respect to $\boldsymbol{\theta}^{c}$.
Since the minimum operation preserves the continuity property of the
original functions, $\mathfrak{C}_{k} (\theta^{c}_k,
\boldsymbol{\theta}_{-k}^{c}) = \min \{
f_{k}^{1}(\theta_{k}^{c},\boldsymbol{\theta}_{-k}^{c}),
f_{k}^{2}(\theta_{k}^{c},\boldsymbol{\theta}_{-k}^{c}) \}$ is thus
continuous in $\boldsymbol{\theta}^{c}$.

\item{The payoff function $\mathfrak{C}_{k} (\theta^{c}_k, \boldsymbol{\theta}_{-k}^{c})$ is quasi-concave in
$\theta_{k}^{c}$.}

Since $w_j\overline{\mathcal C}^{c}_{j}, j\in \mathcal{U}^c$,
$g_{k}^{2}(\theta_{k}^{c},\boldsymbol{\theta}_{-k}^{c})$ and
$f_{k}^{2}(\theta_{k}^{c},\boldsymbol{\theta}_{-k}^{c})$ are the
composition of a linear fractional function (which is quasi-concave
in $\theta_{k}^{c}$) and a logarithm function (which is
non-decreasing), we can conclude that $w_j\overline{\mathcal
C}^{c}_{j}, j\in \mathcal{U}^c$,
$g_{k}^{2}(\theta_{k}^{c},\boldsymbol{\theta}_{-k}^{c})$ and
$f_{k}^{2}(\theta_{k}^{c},\boldsymbol{\theta}_{-k}^{c})$ are
quasi-concave in $\theta_{k}^{c}$. Moreover, since the minimum
operation preserves the quasi-concavity, we can prove that the
payoff function $\mathfrak{C}_k (\theta^{c}_k,
\boldsymbol{\theta}_{-k}^{c})$ is quasi-concave in $\theta_{k}^{c}$.
\end{enumerate}
In addition, since the admissible strategy set of player $k$,
$\mathscr{D}_k$, is a nonempty, convex and compact subset of
Euclidean space $\mathbb R$, we can conclude that at least one NE
exists in the game $\mathscr{G}$ which is the direct result of
\cite{Debreu,Fan,Glicksberg,Saraydar,Liang}.

\subsection{Uniqueness}
To prove the uniqueness of NE, we shall provide the following two
Lemmas before the main proof.
\begin{Lem}[Monotonicity]\label{lem:monotonity}
Define $f^{1}(\theta_{k}^{c},\boldsymbol{\theta_{-k}^{c}})$ to be
{\em monotonic increasing} in $\boldsymbol{\theta^{c}}$, if
$$f^{1}(\theta_{k}^{c},\boldsymbol{\theta_{-k}^{c}})\geq
f^{1}({\theta_{k}^{c}}^{'},{\boldsymbol{\theta_{-k}^{c}}}^{'})$$ for
all $\boldsymbol{\theta^{c}}\succeq
\boldsymbol{{\theta^{c}}}^{'}$\footnote{In this paper, $\succeq$
means componentwise larger or equal.} and the equality holds if and
only if $\boldsymbol{\theta^{c}} = \boldsymbol{{\theta^{c}}}^{'}$.
It can be shown that $g^{1}(\boldsymbol{\theta^{c}})$,
$f^{1}(\boldsymbol{\theta^{c}})$ is monotonic increasing in
$\boldsymbol{\theta^{c}}$. Also we can observe that for given
$\boldsymbol{\theta_{-k}^{c}}$,
$g^{1}(\theta_{k}^{c},\boldsymbol{\theta_{-k}^{c}})$,
$g_{k}^{2}(\theta_{k}^{c})$ and $f_{k}^{1}(\theta_{k}^{c},
\boldsymbol{\theta_{-k}^{c}})$ is monotonic increasing in
$\theta_{k}^{c}$. $f_{k}^{2}(\theta_{k}^{c})$ is monotonic
decreasing in $\theta_{k}^{c}$.
\end{Lem}
\begin{Lem}[Utility Solution]\label{lem:utility_solution}
Given $\boldsymbol{\theta_{-k}^{c}}$, the maximum value of the
player $\mathit{l}$'s utility function
$$\mathfrak{C}_{k}(\theta_{k}^{c},\boldsymbol{\theta_{-k}^{c}}) = \min \left\{ f_{k}^{1}(\theta_{k}^{c},\boldsymbol{\theta_{-k}^{c}}) ,f_{k}^{2}(\theta_{k}^{c}) \right\}$$
exists and is unique. Moreover, we have
$f_{k}^{1}({\theta_{k}^{c}}^{\ast},\boldsymbol{\theta_{-k}^{c}})=f_{k}^{2}({\theta_{k}^{c}}^{\ast})$
at the maximum point ${\theta_{k}^{c}}^{\ast}$.
\end{Lem}
\proof The proof of Lemma \ref{lem:monotonity} is straight-forward and hence omitted due to the page limit. The sketched proof of Lemma \ref{lem:utility_solution} can be summarized as follows.

For any given $\boldsymbol{\theta_{-k}^{c}}$, from Lemma
\ref{lem:monotonity}, we can find that
$f^{1}(\theta_{k}^{c},\boldsymbol{\theta_{-k}^{c}})$ reaches the
minimum value at the point $\theta_{k}^{c}=0$ and the maximum value
at the point $\theta_{k}^{c}=1$. Similarly,
$f_{k}^{2}(\theta_{k}^{c})$ reaches the maximum value at the point
$\theta_{k}^{c}=0$ and the minimum value at the point
$\theta_{k}^{c}=1$. Since
$f_k^{1}(\theta_{k}^{c},\boldsymbol{\theta_{-k}^{c}}) /
f_{k}^{2}(\theta_{k}^{c})$ is continuous and strictly
increasing/decreasing with respect to $\theta_{k}^{c}$, combing with
the fact that $f_{k}^{1}(0,\boldsymbol{\theta_{-k}^{c}})<
f_{k}^{2}(0)$ and
$f_k^{1}(1,\boldsymbol{\theta_{-k}^{c}})>f_{k}^{2}(1)$, we can
conclude that there exists a unique point ${\theta_{k}^{c}}^{\ast}
\in(0,1)$ such that
$f_k^{1}({\theta_{k}^{c}}^{\ast},\boldsymbol{\theta_{-k}^{c}})=
f_{k}^{2}({\theta_{k}^{c}}^{\ast})$, which corresponds to the
maximum value of $\min
\{f^{1}(\theta_{k}^{c},\boldsymbol{\theta_{-k}^{c}}),f_{k}^{2}(\theta_{k}^{c})\}$.
Hence, Lemma \ref{lem:utility_solution} follows.
\endproof

With the well elaborated Lemma \ref{lem:monotonity} and Lemma
\ref{lem:utility_solution}, we shall prove the uniqueness of NE in
our game $\mathscr{G}=[\mathscr{L},\{\mathscr{D}_{k}\},
\{\mathfrak{C}_{k}(\cdot)\} ]$ through the mathematical
contradiction. Suppose there exist two different NEs
$\boldsymbol{\theta}$ and $\boldsymbol{{\theta}^{'}}$. Without loss
of generality, we can category the relationship between
$g^{1}(\boldsymbol{\theta}), g^{1}(\boldsymbol{{\theta}^{'}})$ into
the following three classes:
\begin{enumerate}
\item $g^{1}(\boldsymbol{\theta})<g^{1}(\boldsymbol{{\theta}^{'}})$
, we can conclude that there must exist $\mathit{j}\in\mathscr{L}$ such that
$\theta_{j}<\theta_{j}^{'}$.
\begin{enumerate}
\item   Applying Lemma \ref{lem:monotonity}, we have
$ g_{j}^{2}(\theta_{j})<g_{j}^{2}(\theta_{j}^{'}) $  and
$ f_{j}^{2}(\theta_{j})>f_{j}^{2}(\theta_{j}^{'}) $.
\item Applying Lemma \ref{lem:utility_solution}, we have
$f_{j}^{1}(\boldsymbol{\theta})=f_{j}^{2}(\theta_{j})$ and
$f_{j}^{1}(\boldsymbol{\theta^{'}})=f_{j}^{2}(\theta_{j}^{'})$. From the relation that $ f_{j}^{2}(\theta_{j})>f_{j}^{2}(\theta_{j}^{'}) $, we have $f_{j}^{1}(\boldsymbol{\theta})>f_{j}^{1}(\boldsymbol{\theta^{'}})$.
\label{step:uniqueness_1_b}
\item Since
$f_{j}^{1}(\boldsymbol{\theta})=\min\{g^{1}(\boldsymbol{\theta}),g_{j}^{2}(\theta_{j})\}$,
$f_{j}^{1}(\boldsymbol{\theta^{'}})=\min\{g^{1}(\boldsymbol{\theta}^{'}),g_{j}^{2}(\theta_{j}^{'})\}$,
and $g^{1}(\boldsymbol{\theta})<g^{1}(\boldsymbol{{\theta}^{'}})$, $
g_{j}^{2}(\theta_{j})<g_{j}^{2}(\theta_{j}^{'}) $, we have $
f_{j}^{1} (\boldsymbol{\theta})<f_{j}^{1}(\boldsymbol{\theta^{'}})$.
\label{step:uniqueness_1_c}
\end{enumerate}
Thus, a contradiction between step
\ref{step:uniqueness_1_b} and step \ref{step:uniqueness_1_c} has been found.

\item $g^{1}(\boldsymbol{\theta}) > g^{1}(\boldsymbol{{\theta}^{'}})$, the proof shall follow the same lines as in the previous case with $\boldsymbol{\theta}$ and $\boldsymbol{{\theta}^{'}}$ swapped.

\item $g^{1}(\boldsymbol{\theta}) = g^{1}(\boldsymbol{{\theta}^{'}})$,
since $\boldsymbol{\theta}$ and $\boldsymbol{{\theta}^{'}}$ are two
different NEs, without loss of generality, we assume the
$\mathit{j}$th component of the power allocation ratio vector
satisfy $\theta_{j} < \theta_{j}^{'}$.
\begin{enumerate}
\item Applying Lemma \ref{lem:monotonity}, we have
$ f_{j}^{2}(\theta_{j})>f_{j}^{2}(\theta_{j}^{'}) $.
\label{step:uniqueness_3_a}
\item \label{step:uniqueness_3_b} Let
$g^{1}(\boldsymbol{\theta})=g^{1}(\boldsymbol{{\theta}^{'}})=A$, and
the relation between $f_{j}^{2}(\theta_{j})$ and
$f_{j}^{2}(\theta_{j}^{'})$ can be characterized as follows.
\begin{enumerate}
\item if $A\leq g_{j}^{2}(\theta_{j})<g_{j}^{2}(\theta_{j}^{'})$
\begin{eqnarray*}
\left. \begin{array}{l}
f_{j}^{1}(\boldsymbol{\theta})=\min\{g^{1}(\boldsymbol{\theta}),g_{j}^{2}(\theta_{j})\}\\
f_{j}^{1}(\boldsymbol{\theta^{'}})=\min\{g^{1}(\boldsymbol{\theta^{'}}),g_{j}^{2}(\theta_{j}^{'})\}
\end{array} \right\}  \Rightarrow
\left. \begin{array}{c}
f_{j}^{2}(\theta_{j})=A \\
f_{j}^{2}(\theta_{j}')=A
\end{array} \right\}  \Rightarrow  f_{j}^{2}(\theta_{j})=f_{j}^{2}(\theta_{j}^{'})
\end{eqnarray*}
\item if $g_{j}^{2}(\theta_{j})\leq A<g_{j}^{2}(\theta_{j}^{'})$ or $g_{j}^{2}(\theta_{j})<A\leq g_{j}^{2}(\theta_{j}^{'})$
\begin{eqnarray*}
\left. \begin{array}{l}
f_{j}^{1}(\boldsymbol{\theta})=\min\{g^{1}(\boldsymbol{\theta}),g_{j}^{2}(\theta_{j})\}\\
f_{j}^{1}(\boldsymbol{\theta^{'}})=\min\{g^{1}(\boldsymbol{\theta^{'}}),g_{j}^{2}(\theta_{j}^{'})\}
\end{array} \right\} \Rightarrow
\left. \begin{array}{l}
f_{j}^{2}(\theta_{j})=g_{j}^{2}(\theta_{j})\\
f_{j}^{2}(\theta_{j}')=A
\end{array} \right\} \Rightarrow f_{j}^{2}(\theta_{j})\leq f_{j}^{2}(\theta_{j}^{'})
\end{eqnarray*}
\item if $g_{j}^{2}(\theta_{j})<g_{j}^{2}(\theta_{j}^{'})\leq A$
\begin{eqnarray*}
\left. \begin{array}{l}
f_{j}^{1}(\boldsymbol{\theta})=\min\{g^{1}(\boldsymbol{\theta}),g_{j}^{2}(\theta_{j})\}\\
f_{j}^{1}(\boldsymbol{\theta^{'}})=\min\{g^{1}(\boldsymbol{\theta^{'}}),g_{j}^{2}(\theta_{j}^{'})\}
\end{array} \right\} \Rightarrow
\left. \begin{array}{c}
f_{j}^{2}(\theta_{j})=g_{j}^{2}(\theta_{j})\\
f_{j}^{2}(\theta_{j}')=g_{j}^{2}(\theta_{j}^{'})
\end{array} \right\} \Rightarrow f_{j}^{2}(\theta_{j})<f_{j}^{2}(\theta_{j}^{'})
\end{eqnarray*}
\end{enumerate}
The first ``$\Rightarrow$'' is based on Lemma
\ref{lem:utility_solution} and the second one can be easily verified
through basic mathematical relations. Combining with the above three
cases, we can conclude that $f_{j}^{2}(\theta_{j})\leq
f_{j}^{2}(\theta_{j}^{'})$.
\end{enumerate}
Thus, a contradiction between step \ref{step:uniqueness_3_a} and \ref{step:uniqueness_3_b} has been
found.
\end{enumerate}

In summary, since we can always find a contradiction with the assumption that there exist two
different NEs, we can draw the conclusion that the NE is unique in the non-cooperative game
$\mathscr{G}$.
\endproof

\section{Proof of Theorem \ref{thm:algorithm_convergence}}
\label{pf:thm_algorithm_convergence} To prove the convergence property of the proposed iterative
power allocation algorithm, we shall first establish the following relation about the optimal power
allocation ratio.

\begin{Lem}[Optimal Power Allocation] \label{lem:algorithm_converge}
For each player $k$, given the other players' power allocation
$\boldsymbol{\theta}_{-k}^{c}$, the maximum value of the payoff function could be
determined through
\begin{eqnarray}
\beta_{k}^{c,*}=\max_{j\in\mathcal{U}^c,
l\in\mathcal{U}^p_k}\{\theta_{k}^{c,j}, \theta_{k}^{c,l}\}
\end{eqnarray}
where $\theta_{k}^{c,j}$ are the solutions to
$w_j\overline{\mathcal{C}}^{c}_{j}(\theta_{k}^{c},\boldsymbol{\theta}_{-k}^{c})=f_{k}^{2}(\theta_{k}^{c}),
j\in\mathcal{U}^c$ and $\theta_{k}^{c,l}, l\in\mathcal{U}^p$ is the solution to
$g_{k}^{2}(\theta_{k}^{c})=f_{k}^{2}(\theta_{k}^{c})$, respectively.
\end{Lem}

\proof Applying Lemma \ref{lem:utility_solution}, the maximum value of the payoff
function satisfied the following relation
$f_{k}^{1}(\theta_{k}^{c,*},\boldsymbol{\theta}_{-k}^{c})=f_{k}^{2}(\theta_{k}^{c,*})$
and the optimal power allocation $\beta_{k}^{c,*}$ is one element in the set
$\{\theta_{k}^{c,j}, \theta_{k}^{c,l}\}, j\in\mathcal{U}^c, l\in\mathcal{U}^p_k$. Due to
the monotonic increasing property of the function
$f_{k}^{1}(\theta_{k}^{c},\boldsymbol{\theta}_{-k}^{c})$ with respect to
$\theta_{k}^{c}$, we have Lemma \ref{lem:algorithm_converge}.
\endproof

We now give the rigorous proof of Theorem
\ref{thm:algorithm_convergence} as follows. Without loss of
generality, we assume $\boldsymbol{\theta}^{c,*}$ to be the unique
NE in the non-cooperative game $\mathscr{G}$ and the value of
$g^{1}(\boldsymbol{\theta}^{c,*})$ is equal to be $A^{*}$. For all
$k \in \{1,2,\ldots,K\}$, a direct result of Lemma
\ref{lem:algorithm_converge} shows that
$\theta_{k}^{c,*}=\max\{\alpha_{k}^{c,*},\eta_{k}^{c,*}\}$, where
$\alpha_{k}^{c,*}$ and $\eta_{k}^{c,*}$ are the solutions to the
following equations.
\begin{eqnarray}
f_{k}^{2}(\alpha_{k}^{c}) & = & A^{*} \\
f_{k}^{2}(\eta_{k}^{c}) & = & g_{k}^{2}(\eta_{k}^{c})
\end{eqnarray}

Since $\eta_{k}^{c,*}$ can be locally determined, the remaining is
to find the value of $A^{*}$ through numerical algorithms. A
standard bisection search based argument \cite{Boyd} with provable
convergence property can be applied. Combining with the unique
property of the NE as established in Theorem  \ref{thm:nash_exist},
the optimal power allocation ratio $\boldsymbol{\theta}^{c,*}$ can
be determined. To improve the speed of the convergence, we shall
properly set the initial conditions as follows.
\begin{enumerate}
\item $A_{\min} = g_{k}^{1}(\boldsymbol{\eta^{c,*}}) \leq
g_{k}^{1}(\boldsymbol{\theta^{c,*}}) = A^{*}$.
\item $A_{\max}=g_{k}^{1}\big(\boldsymbol{\theta^{c}}(1)\big) \geq
g_{k}^{1}(\boldsymbol{\theta^{c,*}}) = A^{*}$
\end{enumerate}
where $\boldsymbol{\theta^{c}}(1) = [\theta_{1}^{c}(1), \theta_{2}^{c}(1),\ldots,
\theta_{K}^{c}(1)]$ and $\theta_{k}^{c}(1) = \max \{1-
\frac{2^{A_{min}/w_{l}}-1}{P_{k}L_{lk}R_{k}^{p}}, \eta_{k}^{c ,*} \},
l\in\mathcal{U}^p_k$.

\bibliographystyle{IEEEtran}
%% argument is your BibTeX string definitions and bibliography database(s)
\bibliography{IEEEfull,mybibfile}

\newpage

\end{document}